\documentstyle[preprint,aps]{revtex}
\begin{document}
\preprint{HYUPT-95/20}
\draft
\title{Wilsonian Black Hole Entropy in Quantum Gravity}
\author{Sergei D. Odintsov}
\address{Dept. ECM, Faculty of Physics, \\
Diagonal 647, Universidad de Barcelona, 08028 Barcelona, Spain \\
{\rm and} Tomsk Pedagogical Institute, 634041 Tomsk, Russia}
\author{Yongsung Yoon}
\address{Department of Physics, Hanyang University,
Seoul, 133-791, Korea}
\maketitle
\begin{abstract}
Using Wilsonian procedure (renormalization group improvement)
we discuss the finite quantum corrections to black hole entropy in
renormalizable theories. In this way, the Wilsonian black hole
entropy is found for GUTs (of asymptotically free form, in
particularly) and for the effective theory of conformal
factor aiming to describe quantum gravity in infrared region.
The off-critical regime (where the coupling constants are running)
for effective theory of conformal factor in quantum gravity
(with or without torsion) is explicitly constructed.
The correspondent renormalization group equations for
the effective couplings are found using Schwinger-De Witt
technique for the calculation of the divergences of
fourth order operator.
\end{abstract}

\section{Introduction}

Since the seminal papers by Bekenstein and Hawking \cite{r17}
the exact formula
for intrinsic entropy in black hole solutions of Einstein theory
(so-called Bekenstein-Hawking formula) is quite well-known.
Moreover, as it was found recently \cite{r3}, higher derivative
invariants which naturally appear, in particular,
in the gravitational effective action
\cite{r29} give the additional contributions to the entropy density.
Hence, the Bekenstein-Hawking result becomes only the leading
contribution to the entropy density in such circumstances.

Despite many efforts, we are still very far from the complete
understanding of black hole thermodynamics and in particular
from the understanding of black hole entropy.
(For a different approaches to black hole entropy,
see Refs.\cite{r2,r4,r5,r6,r7,r24,r30}.)

Moreover, the recent studies of quantum corrections
\cite{r2,r4,r6,r8,r9,r10,r11,r12,r13,r14}
(and references therein) to black hole entropy on different
black hole backgrounds show that there appear explicit
divergences in such calculations.
One possibility to understand the origin of these divergences
is connected with the equivalence principle \cite{r31,r32}
(see also Refs.\cite{r10,r33}).
{}From another side, as it has been proposed in Ref.\cite{r2},
these divergent contributions to black hole entropy (which
have been discovered long ago \cite{r4})
have  the ultraviolet nature. Hence, it may be removed by the
renormalization of
the gravitational constants \cite{r2} - what is usual way to
work with ultraviolet divergences in quantum field theory
(for a review of renormalization of quantum field theory in
curved background, see Ref.\cite{r1}).

Using cut-off regularization it has been checked that this
indeed the case at least in the situation with non-extremal
black holes (see Refs.\cite{r2,r8,r9,r10,r11,r12,r13,r14});
That was also checked by
Pauli-Villars regularization in Ref.\cite{r14}. For the extremal
black holes,
where entropy vanishes \cite{r34}, it is not completely clear
that it will be the same. However, it seems \cite{r14}
that possibility to remove all
divergences from black hole entropy via the renormalization of
the gravitational coupling constants exists at least using
Pauli-Villars regularization.

The purpose of this work will be to discuss the quantum
corrections to black
hole entropy using Wilsonian procedure \cite{r18}
(or renormalization group (RG)
improvement \cite{r20}). That gives the way to find the Wilsonian
black hole
entropy beyond one-loop (making summation of the
leading-logarithms of the perturbation theory).
Such procedure is standard now in discussing of the
effective potential in the Standard Model.

The paper is organized as following. In the next section, we
discuss general renormalizable quantum field theory including
scalars, spinors and vectors
(GUT-like theory) on curved background with conical singularity.
Using RG equations for effective couplings,
the RG-improved black hole entropy is found.
In section 3, we consider effective theory for conformal factor
aiming to describe the quantum gravity in infrared limit.
This effective theory for quantum gravity is considered
off-critical regime.
The Wilsonian black hole entropy is found in this theory as well.
In section 4, the conformal sector of quantum gravity with
torsion is considered near IR fixed stable point.
The gravitational and torsion running couplings are found.
Finally, some discussion is given in conclusion.

\section{Quantum Field Theory in Black Hole Background and RG
improved Black Hole Entropy}
We will consider an arbitrary multiplicatively renormalizable
field theory including the scalars $\varphi$, spinors
$\psi$ and vectors $A_{\mu}$ in curved
spacetime which will be chosen to be of black hole
type form, i.e. including the conical singularity. General
arguments tell that such
an action to be multiplicatively
renormalizable should have the form \cite{r1};
\[
L=L_{m} +L_{ext}~,
\]
\[
L_{m}=L_{YM}+\frac{1}{2}
(\nabla_{\mu}\varphi)^{2}+\frac{1}{2}\xi
R\varphi^{2}- \frac{1}{4!}f\varphi^{4}
-\frac{1}{2}m^{2}\varphi^{2} +
i\bar{\psi}(\gamma^{\mu}\nabla_{\mu} -h\varphi) \psi~,
\]
\begin{equation}
L_{ext} = a_{1}R^{2} +a_{2}C_{\mu\nu\alpha\beta}^{2}
 + a_{3}G + \Lambda -\frac{1}{16\pi G} R~,
\label{e1}
\end{equation}
where some gauge group  is supposed to be chosen.
Note that it is quite well-known
(see Ref.\cite{r1} for a review) that the
Lagrangian of the external fields
$L_{ext}$ should be added to $L_{m}$
in order to have the theory to be multiplicatively renormalizable.
As the example of black hole type background we will consider the
Euclidean space which topologically represents the direct
product of the singular surface (horizon surface)
and two-dimensional cone with angle deficit $2\pi(1-\alpha)$.
The example of such type is given by the well-known Rindler space
which represents the infinite mass limit of the
Schwarzschild black hole. The corresponding Euclidean metric is
given by
\begin{equation}
ds^{2}= \rho^{2}d\theta^{2} +
d\rho^{2} +dx_{2}^{2} +dx_{3}^{2}~,
\label{e2}
\end{equation}
where Euclidean time $\theta$ is periodic with period $\beta$,
space coordinates
$x_{2}, x_{3}$ are restricted.
The subspace of constant $x_{2}, x_{3}$
forms a cone with angle deficit $2\pi(1- \alpha)$
(i.e. with conical singularity at $\rho=0$).
However \cite{r2}, in Lorentzian notations where $\rho=0$
describes the black hole horizon there is no singularity.
Hence, for correct Euclidean continuation $\alpha=1$
(i.e, $\beta=\beta_{H}$ as $\alpha =\frac{\beta}{\beta_{H}}$),
where $\beta_{H}= 2\pi$ for the metric (\ref{e2}).

Our purpose in this section will be to calculate the quantum
matter corrections to the
black hole entropy for the theory of type (\ref{e1}) on the above
described black hole background.
Note that since the work by t'Hooft \cite{r4}
(see also Ref.\cite{r24})
where the attempt to calculate the quantum corrections to black
hole entropy has been first done, there
appeared many works where similar calculations have been done
in various contexts and using the different approaches to
entropy \cite{r6,r7,r8,r15} (for a review of discussions,
for different definitions
of black hole entropy, see Ref.\cite{r5}).
It has been checked by the direct calculations that quantum
corrections to black
hole entropy contain  divergences (see, for example,
Refs.\cite{r2,r8,r9,r10,r11,r12,r13,r14}).
It has been proposed in Ref.\cite{r2} that divergences which
appear in the calculation of the quantum corrections to black
hole entropy (more precisely, the divergent part of these
corrections) have the ultraviolet nature
and may be removed by the renormalization of
the gravitational constants. It has been shown by direct
calculations, mainly using the example of free scalar
theory in the frames of cut-off (brick wall) regularization, that
this is indeed the case at least for canonical horizon.
Moreover, the renormalization of gravitational
coupling constants removes
the divergences from the quantum corrections to
black hole entropy even in
the case of using Pauli-Villars regularization as it was shown
in Ref.\cite{r14} recently.
It is clear that if this indeed the case then any other
regularization (which may look less physical in above context)
should work as well while one is working with renormalizable
theories.

Here we suggest to use the dimensional regularization within
scheme of minimal subtraction.
Then, only logarithmic divergences appear in the effective action
calculation.  Moreover, the using of such technique may be useful
presumably in the similar considerations for extremal black holes
where extra cubic divergences in cut-off regularization
have been found \cite{r10,r12}.
These extra divergences in cut-off regularization seems to be
impossible to
be removed by renormalization of gravitational couplings.
This fact led the authors of Ref.\cite{r12} to argue that
thermodynamics
of such black holes is not well-defined.

We will start from the theory (\ref{e1}) on purely gravitational
background of type (\ref{e2}). Then the classical (tree-level)
entropy may be easily defined as following \cite{r3,r16}:
\[
\sigma= \frac{A}{4G}
- \int_{\Sigma} [8\pi (a_{1}+\frac{1}{3}a_{2}+a_{3})R
\]
\begin{equation}
- 4\pi (2a_{2}+4a_{3})R_{\mu\nu}n^{\mu}_{i}n^{\nu}_{i}+
8 \pi (a_{2}+a_{3})R_{\mu\nu\lambda\rho}n^{\mu}_{i}n^{\lambda}_{i}
n^{\nu}_{j}n^{\rho}_{j}]~,
\label{e3}
\end{equation}
where the first term is well-known Bekenstein-Hawking entropy
\cite{r17}, $n^{k}$
are two orthonormal vectors orthogonal to $\Sigma$.
Note that one can use different forms of higher order invariant
corrections to Bekenstein-Hawking entropy (see Ref.\cite{r3} for
details). Notice also that
coupling constants which appear in Eq.(\ref{e3}) are
classical tree- level coupling constants.

The entanglement entropy is defined by the standard relation:
\begin{equation}
\sigma=(\beta\frac{\partial}{\partial \beta} -1)\Gamma~,
\label{e4}
\end{equation}
where $\Gamma$ is one-loop effective action (free energy) on the
background
(\ref{e2}), and after the calculation one has to put
$\beta=\beta_{H}$, i.e. $\alpha=1$ \cite{r2}.

Now one can make the renormalization of the entanglement
entropy via
\begin{equation}
\sigma(G, a_{1}, a_{2}, a_{3})
+ \sigma_{div}=\sigma_{ren}( G^{R}, a_{1}^{R},
a_{2}^{R},a_{3}^{R})~,
\label{e5}
\end{equation}
where on the right-hand side the gravitational
coupling constants $G, a_{1}, a_{2}, a_{3}$ are renormalized
ones (we will drop below the superscript
$R$ for simplicity).
They are connected with the bare couplings in the standard
way \cite{r1} via renormalization of the one-loop
effective action on the regular spacetime.
Schematically, these relations look like
\begin{equation}
G^{-1}_{R}=G^{-1}_{B}+\frac{\tilde{G}m^{2}}{(n-4)}~,~~~
a_{i}^{R} =a_{i}^{B}+\frac{\tilde{a}_{i}}{(n-4)}~,~~~
(i=1,2,3)~,
\label{e6}
\end{equation}
where coefficients $\tilde{G}, \tilde{a}_{i}$ originate from
the well known $a_{2}$- coefficient of Schwinger-De
Witt expansion \cite{r29}.
This coefficient is defined only by quadratic part of action
(\ref{e1}) for all quantum fields, and it was written
explicitly for the fields of spin
$0, \frac{1}{2}, 1, \frac{3}{2}, 2$ in the literature
few hundred times, so
it is not necessary to write it explicitly.
It may be shown in a variety of ways
\cite{r2,r9,r10,r12,r14,r16}
that this renormalization (\ref{e6}) is not influenced by the
presence of the conical singularity.
So, the ultraviolet divergences of the
entropy are given by (\ref{e3}) where instead of
gravitational couplings
one has to substitute their infinite parts
(according to (\ref{e6}) in dimensional regularization, or
one can use cut-off regularization \cite{r2,r4}).
Now, after this discussion of the renormalization of
the black hole entropy
(which definitely works for the canonical horizon)
one can try to get some universal
information from the ultraviolet renormalized entropy.

For that purpose, one may apply the Wilsonian
procedure \cite{r13} (or renormalization group improvement)
which gives the way to make the
summation of leading logarithms of the whole
perturbation series.  This way is based on renormalization
group and it was successfully applied in the study of
the effective potential in Standard Model.

First of all, let us write explicitly the RG equations
for gravitational coupling constants \cite{r1}
(they are induced by the one-loop renormalization (\ref{e6}));
\[
\frac{da_{1}(t)}{dt} =
\frac{1}{(4\pi)^{2}}[\xi(t)- \frac{1}{6}]^{2}
\frac{N_{s}}{2}~,
\]
\[
\frac{da_{2}(t)}{dt} =
\frac{1}{120(4\pi)^{2}}(N_{s}+6N_{f} +12N_{A})~,
\]
\[
\frac{da_{3}(t)}{dt} =
- \frac{1}{360(4\pi)^{2}}(N_{s}+11N_{f}+ 62N_{A})~,
\]
\[
\frac{d\Lambda(t)}{dt}=\frac{m^{4}(t)N_{s}}{2(4\pi)^{2}}~,
\]
\begin{equation}
\frac{d}{dt} \frac{1}{16\pi G(t)}=
- \frac{m^{2}(t)N_{s}}{(4\pi)^{2}}[\xi(t)-\frac{1}{6}]~,
\label{e7}
\end{equation}
where $N_{s}, N_{f}, N_{A}$ are the numbers of real scalars,
spinors and vectors in the theory (\ref{e1}).

Thus, the universal form of RG improved quantum corrections
to entropy looks now as:
\[
\sigma= \frac{A}{4G(t)}
- \int_{\Sigma} [8 \pi \{a_{1}(t)+\frac{1}{3}a_{2}(t)
+ a_{3}(t)\}R
\]
\begin{equation}
- 4\pi\{2a_{2}(t)+4a_{3}(t)\}R_{\mu\nu}n^{\mu}_{i}n^{\nu}_{i}
+8\pi \{a_{2}(t)+a_{3}(t)\}R_{\mu\nu\lambda\rho}n^{\mu}_{i}
n^{\lambda}_{i}n^{\nu}_{j} n^{\rho}_{j}]~.
\label{e8}
\end{equation}
Here, now we have running coupling constants in Eq.(\ref{e8})
instead of classical coupling constants.

For example, in GUT models which are asymptotically free in
all interaction
couplings (for a review, see Ref.\cite{r1}) one has
\[
g^{2}=g^{2}(1+\frac{B^{2}g^{2}t}{(4\pi)^{2}})^{-1}~,
\]
\[
h^{2}=\kappa_{1}g^{2}(t)~,~~~ f(t)=\kappa_{2}g^{2}(t)~,
\]
\[
\xi(t)= \frac{1}{6}+ (\xi- \frac{1}{6})
(1+ \frac{B^{2}g^{2}t}{(4\pi)^{2}})^{b}~,
\]
\begin{equation}
m^{2}(t)= m^{2} (1+ \frac{B^{2}g^{2}t}{(4\pi)^{2}})^{b}~,
\label{e9}
\end{equation}
where $\kappa_{1},\kappa_{2}$ are numerical constants defined
by the specific features of the theory (gauge group, number
of scalars and spinor multiplets),
the constant $b$ may be positive or negative
what depends on the model under consideration.

Then, solving Eq.(\ref{e7}) we will find (see also recent work
Ref.\cite{r19}, where application of the above running coupling
constants for study of quantum corrections to
Newtonian potential has been done):
\[
G(t)=G_{0} \{ 1-
\frac{16\pi N_{s}G_{0}m^{2}(\xi-\frac{1}{6})}{B^{2}g^{2}(2b+1)}
[(1+ \frac{B^{2}g^{2}t}{(4\pi)^{2}}) ^{2b+1} -1] \}^{-1}~,
\]
\[
a_{1}(t)=
a_{1}+ \frac{N_{s}(\xi -\frac{1}{6})^{2}}{2(2b+1)B^{2}g^{2}}
 [(1+ \frac{B^{2}g^{2}t}{(4\pi)^{2}})^{2b+1}-1]~,
\]
\[
a_{2}(t)=a_{2}
+ \frac{t}{120 (4\pi)^{2}}(N_{s} +6N_{f} +12 N_{A})~,
\]
\begin{equation}
a_{3}(t)=a_{3}
- \frac{t}{360 (4\pi)^{2}}(N_{s} +11N_{f} +62 N_{A})~.
\end{equation}

Note that we found only universal quantum corrections to finite
black hole entropy in GUT under consideration making summation
of leading-logarithmic
terms of whole perturbation series, and applying Wilsonian
procedure \cite{r18}.
Of course, there are also finite quantum corrections resulting
from specific features of the black hole background
under consideration.
They were extensively studied recently in
Refs.\cite{r10,r11,r12} in one-loop level.
One may also hope that application of dimensional regularization
may help to solve the problem which appears in extremal
black hole (where cubic divergences in cut-off regularization
have been found) just because powerlike divergences
disappear in dimensional regularization.

Now the natural question appears: what is RG parameter
$t$ in our context?
The choice of this parameter is normally connected with the
presence of the mass parameters in the theory .
In particular, in standard RG $t=\ln\frac{\mu}{\mu_{0}}$
where $\mu, \mu_{0}$ are different mass scales,
in the study of RG improved
potential \cite{r20}
$t=\frac{1}{2}\ln\frac{\varphi^{2}}{\mu^{2}}$
(or $\frac{1}{2}\ln\frac{\varphi^{2}}{\varphi_{0}}$)
where $\varphi$ is scalar field and this choice is not changed
in curved spacetime \cite{r21}.
In the situation under discussion the natural choice is
$t=\ln \beta \mu$ where $\beta$ is temperature and $\mu$ is
an arbitrary mass parameter.
For extremal black hole, the correspondent parameter looks as
$t=\ln\frac{\beta}{\tau_{H}}$ where $\tau_{H}$ is the horizon
radius.

Thus, we found the (beyond one-loop) form of matter quantum
corrections
to black hole entropy which are universal and are caused by
the ultraviolet structure of the theory.

\section{Quantum corrections to Black Hole Entropy in the
Effective Theory for Conformal Factor}
Up to now almost all studies of quantum corrected black hole
entropy have been done for free matter (mainly scalar).
It is very
interesting to understand what happens then in quantum gravity?
Due to the fact that free energy is affected by ultraviolet
divergences, the only consistent way to consider such questions
is in the frames of consistent renormalizable quantum gravity.
In the absence of such theory (for a general review of different
quantum gravity models, see Ref.\cite{r1}),
we suggest to consider the effective
theory of quantum gravity, taking the so-called effective theory
for conformal factor \cite{r22,r23} (which is renormalizable one)
as an example.
We will consider the effective theory for conformal factor
on curved background \cite{r23} which will be chosen of
the black hole type as in previous section.
Note that the effective theory for the conformal factor has been
suggested \cite{r22} to describe the infrared sector of quantum
gravity.

The action of such theory looks as following \cite{r23};
\[
L= -\frac{Q^{2}}{(4\pi)^{2}} \sigma  \Box^{2}\sigma +
\sigma[\xi_{1}R^{\mu \nu} \nabla_{\mu}\nabla_{\nu} +
\xi_{2}R\Box + \xi_{3}(\nabla_{\mu}R) \nabla^{\mu}]\sigma
\]
\[
-\zeta[ 2\tilde{\alpha}(\nabla_{\mu}\sigma)(\nabla^{\mu}\sigma)
\Box\sigma+\tilde{\alpha}^{2}((\nabla_{\mu}\sigma)(\nabla^{\mu}
\sigma))^{2}]
+\frac{\eta_{1}}{\tilde{\alpha}^{2}}e^{2 \tilde{\alpha}\sigma}R
\]
\begin{equation}
+ \eta_{2}R (\nabla_{\mu}\sigma)(\nabla^{\mu}\sigma)
+ \gamma e^{2 \tilde{\alpha} \sigma}
(\nabla_{\mu}\sigma)(\nabla^{\mu}\sigma)
-\frac{\lambda}{\tilde{\alpha}^{2}}e^{4\tilde{\alpha}\sigma} +
\tilde{a_{1}}R^{2}_{\mu\nu}+\tilde{a_{2}}G+\tilde{a_{3}}R^2~,
\label{e11}
\end{equation}
where $\tilde{\alpha}$ is the scaling dimension of scalar field
$\sigma$.
Note that action (\ref{e11}) represents the multiplicatively
renormalizable generalization of the effective theory for the
conformal factor
\cite{r22} on curved background (we suppose the absence of
$\sigma$- linear
terms on classical and quantum level).
Notice also that due to the presence of higher derivative terms,
such a theory may have the problem with unitarity.
As it is effective (not fundamental) theory,
it is not such a big problem. Note also that a mechanism to
ensure unitarity
in such model may exist \cite{r35} in the original theory for
conformal factor.
One starts from the conformal anomaly for free conformally
invariant matter:
\begin{equation}
T=b(C^{2}_{\mu \nu \alpha \beta} +\frac{2}{3}\Box R)
+b^{'}G +b^{''}\Box R~.
\label{e12}
\end{equation}
Working in the parameterization
\begin{equation}
g_{\mu \nu}= e^{2\sigma(x)}\bar{g}_{\mu\nu}~,
\label{e13}
\end{equation}
where $\sigma(x)$ is the conformal factor which is supposed to
be mainly dominant on quantum level in infrared sector of
quantum gravity \cite{r22}
and $\bar{g}_{\mu \nu}$ is a fixed fiducial metric, one can
integrate Eq.(\ref{e12}) in the parameterization (\ref{e13})
(see Ref.\cite{r1} for example) and obtain the anomaly induced
effective action.
After addition to this anomaly induced effective action the
standard Einstein
action (with gravitational constant $\kappa$ and cosmological
constant $\Lambda$) in parameterization (\ref{e13}) one obtains
the action of the from (\ref{e11}) with:
\[
\frac{Q^{2}}{(4\pi)^{2}}= 2b+3b^{''}~,~~~
\zeta= 2b+2b^{'}+3b^{''}~,~~~ \gamma=\frac{3}{\kappa}~,~~~
\lambda=\frac{\Lambda}{\kappa}~,
\]
\begin{equation}
\xi_{1}= 2(\zeta-\frac{Q^{2}}{(4\pi)^{2}})~, ~~~
\xi_{2}=-\zeta + \frac{2}{3} \frac{Q^{2}}{(4\pi)^{2}}~,~~~
\xi_{3}=-\frac{Q^{2}}{3(4\pi)^{2}}~.
\label{e14}
\end{equation}
In order to have such a theory to be multiplicatively
renormalizable, one has to add few more terms to such
a action with arbitrary coupling constants ($R^{2}$ -
terms and terms with
$\eta_{1}, \eta_{2}$).
That is done in the writing of the action (\ref{e11}).

Now, choosing as it was said before, the metric $g_{\mu\nu}$
in the black hole type form (like (\ref{e2})) and
background $\sigma$- field
to be zero, one can easily write the tree level black hole
entropy as:
\begin{equation}
\sigma= -4\pi \frac{A}{\tilde{\alpha}^{2}}\eta_{1} -
\int_{\Sigma}[ 8\pi (\tilde{a_{3}} + \tilde{a_{2}}) R
+ 4\pi(\tilde{a_{1}} -4\tilde{a_{2}}) R_{\mu \nu}
n^{\mu}_{i}n^{\nu}_{i} +
8\pi \tilde{a_{2}}R_{\mu \nu \lambda \rho} n^{\mu}_{i}
n^{\lambda}_{i} n^{\nu}_{j} n^{\rho}_{j}] ~.
\label{e15}
\end{equation}
We suppose that in the IR sector of quantum gravity (which may be
relevant for black hole physics) the main contribution of quantum
gravity to entropy comes from the conformal sector, i.e.
rest gravitational modes are frozen as in the original paper
\cite{r22}.

Using the results of Refs.\cite{r23,r25}, we may find
the running gravitational couplings which present in the expression
for the entropy.
The  correspondent expressions are given by
an asymptotically free solution for $\zeta(t)$ in infrared
\cite{r25};
\begin{equation}
\zeta(t)=- \frac{4 b^{'2} (4\pi)^{2}}{5\tilde{\alpha}^{2}t}~,~~~
Q^{2}(t)=(4\pi)^{2}[ \zeta(t)-2b^{'}]~,~~~
{\rm where}~~|t|~~{\rm is~~big}, ~~~t<0~.
\label{e16}
\end{equation}
Using the running couplings (\ref{e16}) and the result of
Ref.\cite{r23}
one can easily estimate the gravitational couplings in (\ref{e15})
in IR sector ($t \rightarrow - \infty$);
\begin{equation}
\tilde{a_{1}}(t)= \tilde{a_{1}}- \frac{2t}{15(4\pi)^{2}}~,~~~
\tilde{a_{2}}(t)= \tilde{a_{2}}+ \frac{t}{90(4\pi)^{2}}~,~~~
\tilde{a_{3}}(t)= \tilde{a_{3}}+ \frac{2t}{45(4\pi)^{2}}~.
\label{e17}
\end{equation}
Note that only leading terms are kept in the expressions (\ref{e17}).
In the similar way one can obtains the running coupling $\eta_{1}(t)$
which plays the role of gravitational coupling in Eq.(\ref{e15}).
It is given by (for simplicity $\eta_{1}(0)$ is chosen to be zero);
\begin{equation}
\eta_{1}(t) \simeq (-t)^{\frac{2}{5}}
e^{(2-2\alpha + \frac{2\tilde{\alpha}^{2}}{Q^{2}(0)})t}~.
\label{e18}
\end{equation}
One sees that even choosing $\eta_{1}(0)=0$ on classical level,
we have this term induced by quantum corrections
which leads also to its appearance in black hole entropy.
Thus, substituting instead of tree level couplings $\eta_{1},
\tilde{a_{1}}, \tilde{a_{2}}, \tilde{a_{3}}$ in (\ref{e15}),
the correspondent running couplings found before we immediately
obtain the improved expression for the black hole entropy from
quantum effects of infrared gravity.
The choice of RG parameter $t$ was already discussed at the end of
the previous section.

In similar way one can find the universal form for the quantum
corrections to black hole entropy in other renormalizable models
of quantum gravity.
One example may be $R^{2}$-gravity \cite{r1} (forgetting again
about unitarity problem) where the gravitational running coupling
constants are known \cite{r1}
(for a recent discussion, see also Ref.\cite{r19}).

\section{Conformal Sector of Quantum Gravity with Torsion near IR
Fixed Point}

As we discussed in the previous sections, the running of
gravitational
couplings may be relevant for the study of quantum corrections
to black hole entropy.
Let us discuss now the running couplings in the effective theory
of conformal factor in quantum gravity with torsion
(see Ref.\cite{r26} where such a theory has been constructed).
Due to technical problems which result in the appearance of many
new terms in conformal anomaly for theory with torsion,
we will consider as in Ref.\cite{r26} the flat background
(i.e. $g_{\mu\nu}=e^{2\sigma}\eta_{\mu\nu}$)
and anti-symmetric part of torsion
$S_{\mu}$ to be non-zero only.
Then the correspondent Lagrangian is given by;
\[
L= -\frac{Q^2}{(4\pi)^{2}}(\Box \sigma)^{2} -
\zeta[2\tilde{\alpha}(\partial_{\mu}\sigma)^{2}\Box\sigma+
\tilde{\alpha}^{2}( \partial_{\mu} \sigma)^{4}] +
\gamma e^{2\tilde{\alpha}\sigma}(\partial_{\mu}\sigma)^{2}
\]
\begin{equation}
-\frac{\lambda}{\tilde{\alpha}^{2}}e^{4 \tilde{\alpha}\sigma}+
b_{1}S^{2} (\partial_{\mu} \sigma)^{2}+
b_{2}(S^{\mu}\partial_{\mu} \sigma)^{2} +
\frac{h}{2\kappa\tilde{\alpha}^{2}}e^{2\tilde{\alpha}\sigma}
S^{2}+ \eta S^{4}~.
\label{e19}
\end{equation}
All the details of the construction of this Lagrangian,
which consists basically of three pieces;
first, anomaly-induced Lagrangian \cite{r1}, second, the Einstein
theory action with torsion and third, the last term
in (\ref{e19}) introduced for the theory to be renormalizable,
  are given in Ref.\cite{r26}.

Note that running of coupling constants $Q^{2}, \zeta(t),
\gamma(t), \lambda(t)$
near IR stable fixed point $(\zeta=0)$ has been given in
Ref.\cite{r25} (see also Eq.(\ref{e16}) of
the previous section).
Scaling dimensions of the theory in IR fixed point have been
given in Ref.\cite{r22}.

The calculation of the $\beta$-functions in the theory
(\ref{e19}) at $\zeta=0$ has been done in Ref.\cite{r22} and at
$\zeta\ne 0$
(off-critical regime) in Ref.\cite{r25} for
the case of zero torsion.
Now, using the standard algorithm for the calculation of the
divergences of 4-th derivative scalar operator
(see Ref.\cite{r23}), we may get, in the sector connected
with the torsion, the
corresponding $\beta$-functions (we don't give the details
of this background field method calculation,
for an introduction to this method, see for example Ref.\cite{r1});

The $\zeta$ and torsion dependent terms of the 2nd heat kernel
coefficient $a_{2}$ (see Refs.\cite{r29,r36} for definition)
is given by
\[
a_{2}=e^{2\tilde{\alpha}\sigma}S^{2}[\frac{h}{\kappa Q^{2}}
+ \frac{\gamma (4\pi)^{2}}{Q^{4}}(b_{1}+\frac{1}{4}b_{2})]
+S^{4}\frac{(4\pi)^{2}}{2Q^{4}}
[b_{1}^{2}+\frac{1}{2}b_{1}b_{2}+\frac{1}{8}b_{2}^{2}]
\]
\begin{equation}
+S^{2}(\partial\sigma)^{2}
\frac{\zeta\tilde{\alpha}^{2}}{3Q^{2}}
[(9b_{1}+2b_{2}) - 2\zeta(6b_{1}+b_{2})] +
(S^{\mu}\partial_{\mu}\sigma)^{2}
\frac{\zeta\tilde{\alpha}^{2}}{3Q^{2}}b_{2}(1-4\zeta)
\label{e0}
\end{equation}
Using this $a_{2}$ coefficient, we have found the following
RG equations;
\[
\frac{dh(t)}{dt}=
-\frac{3 (4\pi)^{2}\tilde{\alpha}^{2}\zeta(t)}{Q^{4}(t)}h(t)
+\frac{6 (4\pi)^{2}\tilde{\alpha}^{2}
\gamma(t)}{Q^{4}(t)}(b_{1}(t)+\frac{1}{4}b_{2}(t))~,
\]
\[
\frac{db_{1}(t)}{dt}=\frac{\zeta(t)
\tilde{\alpha}^{2}}{3Q^{2}(t)}(9b_{1}(t)
+2b_{2}(t)) +O(\zeta^{2}(t))~,
\]
\[
\frac{db_{2}(t)}{dt}=
\frac{\zeta(t)\tilde{\alpha}^{2}}{3Q^{2}(t)}b_{2}(t)
+O(\zeta^{2}(t))~,
\]
\begin{equation}
\frac{d\eta(t)}{dt}=
\frac{(4\pi)^{2}}{2Q^{4}(t)}[b_{1}^{2}(t) +
\frac{1}{2}b_{1}(t) b_{2}(t)+\frac{1}{8}b_{2}^{2}(t)]~.
\end{equation}
A set of solution for the above RG equations in the limit
$t \rightarrow -\infty$ is
\[
b_{1}(t) = -\frac{1}{4}b_{2}(t)~,
\]
\[
b_{2}(t) \simeq c (-t)^{-\frac{Q_{0}^{2}}{15(4\pi)^{2}}}~,
\]
\[
\eta(t) \simeq
-\frac{c^{2}(4\pi)^{2}}{15Q_{0}^{4}
(1-\frac{2Q_{0}^{2}}{15(4\pi)^{2}})}
(-t)^{1-\frac{2Q_{0}^{2}}{15(4\pi)^{2}}} + c^{'}~,
\]
\begin{equation}
h(t) \simeq c^{''}(-t)^{\frac{3}{5}}~.
\end{equation}
where $c, c^{'}$ and $c^{''}$ are the integration constants.
Note that in the solution for $\eta(t)$ one has to suppose that
the number of matter fields is large enough in order that
the condition $\frac{2Q_{0}^{2}}{15(4\pi)^{2}} > 1$
would hold. Otherwise, $\eta(t)$ would be the increasing
function of $t$.  Notice also that, even though $h(t)$ is a
growing function, the coupling of
$e^{2\tilde{\alpha}\sigma}S^{2}$ term in the Lagrangian
(\ref{e19}) is $h(t)\gamma(t) \simeq c^{''}(-t)^{\frac{2}{5}}
e^{t(2-2\tilde{\alpha}+\frac{2\tilde{\alpha}^{2}}{Q_{0}^{2}})}$,
which is a decreasing function in far infrared provided
$(2-2\tilde{\alpha}+\frac{2\tilde{\alpha}^{2}}{Q_{0}^{2}}) > 0$.
This condition is necessary also to ensure that $\gamma(t)$ is a
decreasing function in far infrared.
Therefore, the decreasing of all running couplings in far infrared
$t \rightarrow -\infty$ for the Lagrangian (\ref{e19}) is
guaranteed if the following conditions hold;
\[
\frac{2Q_{0}^{2}}{15(4\pi)^{2}} > 1~~~{\rm and}~~~
(2-2\tilde{\alpha}+\frac{2\tilde{\alpha}^{2}}{Q_{0}^{2}}) > 0~.
\]

Thus, we constructed the off-critical regime in effective
theory for conformal factor in quantum gravity with torsion.
However, to apply these results to calculation of quantum
corrections to black hole entropy one has to consider
the original theory on curved background.
That requires the addition
of many more terms to the original Lagrangian. In addition, one
has to construct the entropy for the effective gravitational
Lagrangian which includes many torsion terms. These questions
will be discussed in other place.

\section{Discussion}

We discussed the RG improving procedure to obtain the
universal quantum corrections to the entropy on the black
hole background. The following renormalizable theories
have been considered: an arbitrary renormalizable GUT and
the effective model for quantum gravity. As usually, Wilsonian
procedure gives the universal way to get some information
about quantum properties of the system under consideration.

The (beyond one-loop) Wilsonian black hole entropy which has
been discussed in this work has its origin in the logarithmic
divergences of the whole perturbation series. It gives only
universal part of necessary information. There are also
quantum corrections which depend very much from the choice
of black hole background under consideration.
These finite corrections
are not universal and should be calculated directly using
specific black hole background.

We also constructed off-critical regime in the effective
theory for conformal factor in quantum gravity (with torsion).
In particularly, the running couplings have been defined.
They may be useful not only for calculation of black hole
entropy but in other cosmological applications,
like quantum corrections to Newtonian potential or
influence of the running gravitational couplings in
galaxy formation processes. \\~\\ \noindent
{\it Acknowledgments:}
SDO would like to thank I. Antoniadis and E. Mottola for
helpful discussions.
This work was supported in part by
MEC, DGICYT (Spain), CIRIT (Generalitat de Catalunya),
Korea Science and Engineering Foundation, and
the Ministry of Education (Korea) through BSRI-2441.

\end{document}